\documentclass[twocolumn,showpacs,prb]{revtex4}
\usepackage{graphicx}
\usepackage{dcolumn}
\usepackage{amsmath}
\usepackage{latexsym}
\usepackage{longtable}
\begin{document}
\title{Temperature dependent electrical resistivity of a single strand of ferromagnetic single crystalline nanowire}
\author{M. Venkata Kamalakar and  A. K. Raychaudhuri}
\affiliation{DST Unit for NanoSciences, Department of Material Science, S.N.Bose National Centre for Basic Sciences, Block JD, Sector III, Salt Lake, Kolkata - 700 098, India.}
\author{Xueyong Wei, Jason Teng and Philip D. Prewett}
\affiliation{School of Mechanical Engineering, University of Birmingham, United Kingdom.}
\date{\today}
\begin{abstract}

\noindent
We have measured the electrical resistivity of a single strand of a ferromagnetic  Ni nanowire of diameter 55 nm using a 4-probe method in the temperature range 3 K-300 K. The wire used is chemically pure and is  a high quality oriented single crystalline sample in which the temperature independent residual resistivity is determined predominantly by surface scattering. Precise evaluation of the temperature dependent resistivity ($\rho$) allowed us to identify quantitatively the electron-phonon contribution (characterized by a Debye temperature $\theta_R$) as well as the spin-wave contribution which is significantly suppressed upon size reduction. 
\end{abstract}

\maketitle
\noindent

The resistivity ($\rho$) of magnetic nanowires is of immense interest both from scientific and technological points of view. The reduction in size can lead to both qualitative and quantitative changes in comparison to electrical transport in wires of much larger dimensions. While a lot of work  has been reported on the magnetism in the ferromagnetic nanowires, the electrical transport (in particular, the temperature dependent part) is largely unexplored. In particular, there has been lack of extensive data on high quality single crystalline ferromagnetic nanowires over a large temperature range that allows a quantitative evaluation of the resistivity data and the contributions from different sources of electron scattering. Electrical and thermal measurements on single nickel nanowire with lateral dimension 100 nm $\times$ 180 nm, have been reported\cite{ou}. However these measurements were carried out on a polycrystalline wire with much higher resistivity which makes the relative contribution of the temperature dependent terms much weaker and the evaluation of magnetic contribution in particular becomes difficult. We present, in this letter, a concise study of electrical transport over the temperature range 3 K to 300 K in an oriented ((220)), single crystalline cylindrical nickel nanowire of much smaller diameter (55 nm). The single crystallinity is important because we can avoid the grain boundary contribution to scattering which often leads to high residual resistivity and thus masks other important phenomena at low temperatures. The results and the analysis presented below allow us to clearly separate out all the contributions to the resistivity quantitatively. The results show that in such single crystalline  nanowires the electrons can reach a mean-free path of the order of $1.1$ times the diameter. The size reduction leads to a reduction of the Debye temperature ($\theta_R$) by nearly $30\%$ and substantial reduction of the magnetic contribution to electrical resistivity, which validates the more complex analysis carried out in case of parallel-nanowire arrays[10].

\noindent
The nanowires used in this experiment were prepared by pulsed potentiostatic electrodeposition of Ni inside commercially\cite{synkera} obtained nanoporous anodic alumina templates of thickness $\sim$ 56 $\mu$m with average pore diameter of $\sim$55 nm . The deposition was carried out in a bath containing a 300 g/l $NiSO_4 .6H_2O$, 45 g/l $NiCl_2 .6H_2O$, 45 g/l $H_3BO_3$ electrolyte with the working electrode (a 200 nm silver layer evaporated on one side of the template) at a pulse potential of -1 V with respect to the reference electrode (Saturated Calomel), with 80\% duty cycle and a pulse period of 1 second. The nanowires formed inside the cylindrical pores of the templates were characterized by X-ray Diffraction (XRD) and Transmission Electron Microscopy (TEM) and found to be single crystalline FCC with a preferential growth direction along (220) direction. The representative TEM data and the XRD data are shown in Fig. 1 (a) and (b) respectively. The wires have average diameter of $\approx$ 55 nm as measured from the TEM data. The high resolution TEM data show absence of grain boundary over the length of the wire. The wires so grown are ferromagnetic, as established by the M-H curves. The typical M-H curves at 300 K shown in Fig. 1 were taken at H parallel and perpendicular to the wire long axis by retaining the Ni wires in the alumina template. The M-H curves reveal the highly anisotropic magnetic nature of the array of wire with coercivities of 768 Gauss and 188 Gauss for parallel and perpendicular configurations of the wire axis with the magnetic field respectively. 
\begin{figure}[t]
\begin{center}
\includegraphics[width=8cm,height=5.74cm]{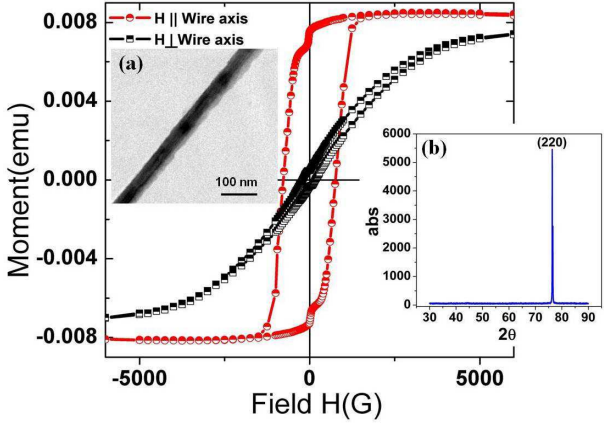}
\end{center}
\caption{M-H curves of the nanowire arrays with measuring field (H) parallel and perpendicular to the wire axis. Inset (a) TEM of 55 nm oriented Ni nanowire. Inset (b) XRD of the sample.}
\label{Fig. 1}
\end{figure}
\noindent
For electrical  measurements on a single nanowire, the wires  were removed from the template by dissolving the latter in a 6M NaOH solution and subsequently washing with millipore water several times. One to two drops of the suspension containing the Ni nanowires were sprayed in the middle portion of a silicon substrate (with 300 nm oxide layer) containing gold contact pads of thickness 500 nm which we made by UV lithography. A relatively long nanowire was chosen under the electron  microscope and the  probes ($\sim$750 nm wide and 300 nm thick) were attached to the nanowire connecting them to the bigger gold contact pads by focused ion beam (FIB) assisted platinum deposition. The inset of Fig. 2 shows the typical image of a nickel nanowire of diameter 55 nm with 5 probes on it. 4-probes were used for the measurement of the resistance of the single strand of the nanowire. To avoid electromigration damage we used an AC signal with a low current amplitude of $10^{-6}A$ (current density $\approx 4\times 10^{8}A.m^{-2}$) with a frequency of 174.73 Hz. The resistance (typically 20-30 Ohms) was measured using a phase sensitive detection scheme with a resolution of 10 ppm. Fig. 2 shows the electrical resistivity of the single Ni nanowire  measured from 3 K-300 K as compared to a 50 $\mu$m thick nickel wire which is the "bulk" reference. It should be noted that this is the first report of electrical measurement on an oriented single crystalline nickel nanowire of this dimension.  It is evident from Fig. 2 that the residual resistivity ratio (RRR) of the single nanowire ($\approx 2.3$) is much less than that of the bulk wire ($\approx 312$). The residual resistivity of the nanowire (even though lower in diameter) is much less than reported earlier\cite{ou} because of its better crystallinity.
\begin{figure}[t]
\begin{center}
\includegraphics[width=8cm,height=6.3cm]{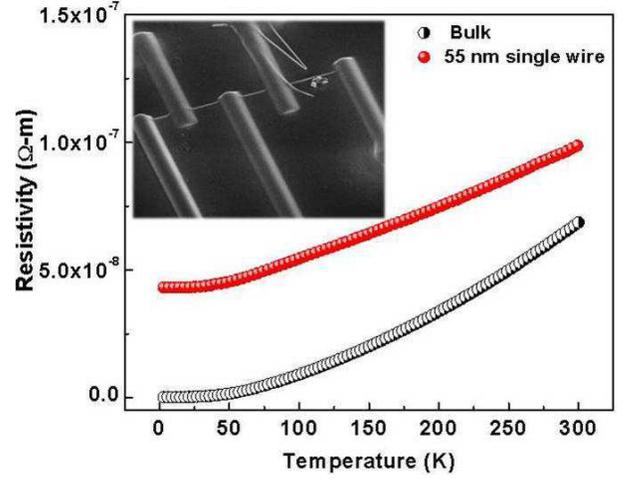}
\end{center}
\caption{Resistivity of the nanowire as compared with the bulk wire. The inset shows the Scanning Electron Microscope image of the nanowire connected to 5 Pt probes made using FIB assisted platinum deposition.}
\label{Fig. 2}
\end{figure}
In general, the resistivity ($\rho$) of a ferromagnetic metal is composed of  the residual resistivity ($\rho_{0}$),  resistivity due to electron-phonon (lattice) interactions ($\rho_{L}$) and the   resistivity due to electron-spin scattering ($\rho_{M}$). By using Matthiessen's rule\cite{ziman} we can write:

In general, the resistivity ($\rho$) of a ferromagnetic metal is composed of  the residual resistivity ($\rho_{0}$),  resistivity due to electron-phonon (lattice) interactions ($\rho_{L}$) and the   resistivity due to electron-spin scattering ($\rho_{M}$). By using Matthiessen's rule\cite{ziman} we can write:

\begin{eqnarray}
\rho&=&\rho_{0}+\rho_{L}+\rho_{M}
\end{eqnarray}

In ferromagnetic metals $\rho_{L}$ is well described by the following Bloch-Wilson(BW) formula ($n=3$)\cite{ziman}.
\begin{eqnarray}
\rho_L&=&\alpha_{el-ph}{(\frac{T}{\theta_R})}^n\int_0^{\frac{\theta_R}{T}}\frac{x^n dx}{(e^x-1)(1-e^{-x})}
\end{eqnarray}
While at low temperature ($T \le 15$ K) the magnetic part behaves like $\rho_{M}= B T^2$ as described by Mannari\cite{mannari, white}.
\begin{eqnarray}
\rho_{M}=BT^2 ;B=\frac{3\pi^5 S \hbar}{16e^{2}k_F}(\frac{\mu}{m})^{2}\frac {(k_{B}T NJ(0))^{2}} {E_{F}^4}
\end{eqnarray}

where S (=1/2) is spin of the conduction electron, $\mu$ is the effective magnon mass, $m$ is the electron mass. For Ni, the ratio $\mu/m$ is $\approx$ 38, $E_F$ is the Fermi energy, $k_{F}$ = Fermi wave vector, and $NJ(0)$ is the strength of the s-d interaction in the long wavelength limit ($\approx$ 0.48 eV for Ni), N being the number of  spins. For Ni the above relation gives $B\approx 1.1\times 10^{-13} \Omega.mK^{-2}$ which matches very well with that  obtained experimentally in bulk Ni.

Beyond 15 K, $\rho_{M}$ has a complicated temperature dependence\cite{goodings}. For 15 K$<T<$100 K, the temperature dependence arises mainly from the lattice contribution $\rho_{L}$ when compared with $\rho_{M}$. Nevertheless at high temperatures ($T >$150 K), $\rho_{M}$  becomes significant again.
\noindent
The data for temperature T$ < $15 K, was  fit with Eq. (1) with the magnetic contribution $\rho_{M}=B T^2$. The  constant $B$ for the single nanowire was  found to be $1.01\times 10^{-13}\Omega.mK^{-2}$. Similar analysis of the reference bulk wire data gives $B=1.529 \times 10^{-13}\Omega.mK^{-2}$ which is close to the values reported in bulk nickel\cite{mannari, white}. The data for the nanowire thus show a significant reduction in the magnetic contribution at low T as shown in the inset of Fig. 3(b).

In the  temperature range 15 K-100 K we could use the BW formula\cite{wilson} (Eq.(1)) to fit the behavior of $\rho(T)$ neglecting $\rho_{M}$. We obtained a reasonably good fit (Fig. 3(a)) with an error of less than 0.3$\%$ shown as an inset in Fig. 3(a). The data thus can be described by a single parameter, namely the Debye temperature ($\theta_{R}$), which as obtained from our analysis is 345 K. This is less than value of 471 K (matches with those obtained from specific heat measurements\cite{ziman}) obtained for the reference bulk wire. The resistivity data thus allow an unambiguous determination of the relevant Debye temperature provided the analysis is carried out in the proper temperature range where electron-phonon interaction is the dominant. The suppression in $\theta_{R}$ can be attributed to an increase in relative surface to volume ratio. As more surface atoms have missing bonds giving rise to larger vibrational amplitudes with low frequencies and hence a lower effective Debye temperature. Such decrease in Debye temperature is also reported in case of gold thin films\cite{gold} and also nanowires of FCC structured nanowires of Silver and Copper\cite{aveek}. Also, the $\theta_{R}$ for different metal nanowires of FCC structure is seen to follow a characteristic trend with diameter of the nanowires\cite{venkat}.
\begin{figure}[t]
\begin{center}
\includegraphics[width=8cm,height=7.1cm]{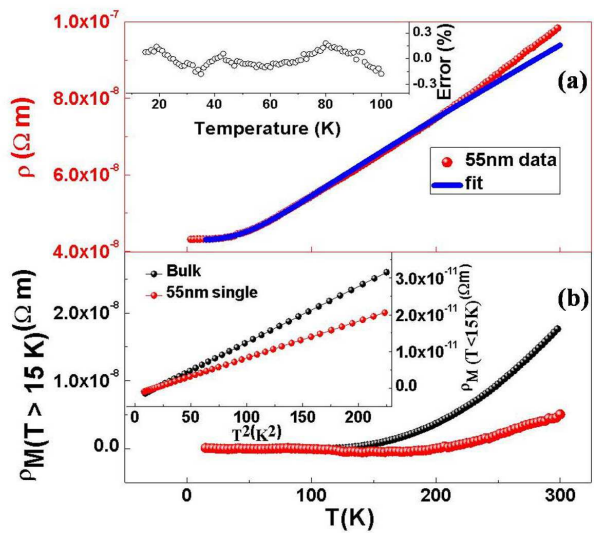}
\end{center}
\caption{(a)Electrical resistivity of the single Ni nanowire along with the fit to the Bloch-Wilson relation  upto 100 K. The inset shows the fit error (\%). (b) Magnetic part of resistivity $\rho_{M}=\rho(T)-\rho_{0}-\rho_{L}$ for T $> $15 K.  The inset shows the magnetic part of resistivity with quadratic temperature dependence  for T $<$ 15 K.}
\label{Fig. 3}
\end{figure}
\noindent
\noindent
The magnetic contribution ($\rho_M$) for $T > 15 K$, is estimated by subtracting out the extrapolated $\rho_L$ and $\rho_0$ from the total measured $\rho$ as shown in Fig. 3(b). The reference bulk data are also shown for comparison.  $\rho_{M}$ has a negligible value in the temperature range 15 K -150 K. For $T > 150$ K also there is a significant suppression of the magnetic scattering in comparison to that seen in the reference bulk wire. We can clearly see an effective decrease in the magnetic contribution to the resistivity as one goes from bulk to nanowires. The curie temperature is often seen to decrease with decrease in size and so as magnetization and hence the interaction strength $N J(0)$ is reduced upon size reduction. We suggest a likely mechanism for this reduction in case of nanowire, in which  the contribution of the disordered spins at the surface become increasingly important. These disordered  surface spins may not support long wavelength propagating spinwaves that are needed for the temperature dependent resistivity of the type that gives $\rho_M$. This will reduce $N$ effectively and hence $N J(0)$ is reduced lowering the magnetic resistivity. We note that  the suppression of the Debye temperature and magnetic resistivity are observed even if we increase the upper limit of temperature $>$100 K in our analysis, though in such a case the fit errors also increase considerably giving rise to large unphysical values of $\theta_{R}$. In our analysis, we took the upper limit of temperature as 100 K for optimum fitting conditions (i.e., to include as much as data with minimum fit percentage error), for best results.

To evaluate the effect of surface scattering, we have used the surface scattering model \cite{dingle} given below for wires of diameter $d << l$, $l$ being the electron mean free path in the bulk sample  
\begin{eqnarray}
\frac{\rho_{0}}{\rho_{0}^{bulk}}&=&\frac{(1-p)}{(1+p)} \frac{l}{d}
\end{eqnarray}

Where $\rho_{0}$  and $\rho_{0}^{bulk}$ are the residual resistivities of the nanowire  and that of the bulk metal respectively. We obtained $p$ = 0.018, the specularity coefficient, which is the fraction of electrons getting elastically scattered from the wire boundary ($p$=1 for completely specular surface and for diffused scattering $p\rightarrow$0).
Using $p$ = 0.018, we estimated the mean free path at 4.2 K given by\cite{sondheimer}
\begin{eqnarray}
l_{NW}&=&\frac{1+p}{1-p} d
\end{eqnarray}

The mean free path $l_{NW}\approx 1.037d$ implies that the mean free path is determined predominantly by surface scattering and the electrons do not suffer significant scattering within the volume of the nanowire because of high purity and fewer defects.

In conclusion, we have reported the electrical resistivity in a single crystalline and oriented Ni nanowire. The resistivity in the temperature range 3 K - 300 K was measured by a  four-probe method with FIB-deposited Pt contacts. The single crystalline nature of the wire ensures that the temperature independent residual resistivity is determined mainly by the surface scattering. We find that the decrease in diameter significantly decreases the Debye temperature ($\theta_{R}$). The magnetic part of the resistivity is remarkably supressed in the case of the nanowire. The single crystalline Ni nanowires with the temperature dependent resistivity being almost linear for $T > 100 K$ and predominantly determined by a single parameter $\theta_{R}$, might be excellent temperature sensors with nanometric dimensions and thus, with very rapid response time.

The authors thank the Unit for NanoScience and Technology, IACS, Kolkata for TEM support. This work is financially supported by the DST, Govt. of India and CSIR, Govt. of India. The Royal Society is thanked for sponsored collaboration between S.N.B.N.C.B.S. and University of Birmingham. One of the authors (M.V.K) acknowledges CSIR, Govt. of India for fellowship.

\end{document}